# Last Query Transformer RNN for knowledge tracing

**SeungKee Jeon**

Samsung Electronics

123use321@gmail.com

## Abstract

This paper presents an efficient model to predict a student's answer correctness given his past learning activities. Basically, I use both transformer encoder and RNN to deal with time series input. The novel point of the model is that it only uses the last input as query in transformer encoder, instead of all sequence, which makes QK matrix multiplication in transformer Encoder to have O(L) time complexity, instead of $O(L^2)$. It allows the model to input longer sequence. Using this model I achieved the 1st place in the 'Riiid! Answer Correctness Prediction' competition hosted on kaggle.

## Introduction

Knowledge tracing, modeling a student's knowlege state based on the history of their learning activities, is one of the rapidly emerging area in AI research. Riiid Labs, an AI education solutions provider, released EdNet[1] which is the world's largest open database for AI education containing more than 100 milion student interactions. Using the dataset, they hosted a competition named 'Riiid! Answer Correctness Prediction'[2] on kaggle, challenging you to create the algorithms for predicting a student's answer correctness given his past learning activities. Submissions were evaluated on area under the ROC curve between the predicted probability and the observed target. This paper presents my winning model to the competition.

## Related Works

Deep learning based models have been shown great performance in knoledge tracing. Recurrent Neural Network (RNN) has been often used[3,4,5], since it well fits the time-series charateristic of the problem. Transformer[6] also has been actively used[7,8,9,10], because its self-attention mechanism has been proved effective for sequential prediction tasks.

## Methodology

### Input feature

For input, I use 5 features, which are question id, question part, answer correctness, current question elapsed time, and timestamp difference. 'Current question elapsed time ' feature means the time elapsed for user to solve the current question. It is a transformed feature from the original database feature 'prior question elapsed time'. 'Timestamp difference' feature indicates the difference from the past question timestamp to the current question timestamp, and it is clipped by maximum value, 3 day. Categorical embedding is used for first three features, and continuous embedding is used for last two continuous features. And then features were merged by DNN to make k-th input feature $I_k$.

### Validation Strategy

Among the given user IDs, first 95% of user IDs are used for training and remaining 5% are used for validation.

### Model

The model structure consists of one custom transformer encoder, one LSTM layer over it, and final DNN to predict correctness. The main idea of using transformer encoder is to make model to learn relationship between questions, and using LSTM over it is to make the model to learn the sequential characteristic of the problem, giving more importance to recent activities. You can see the structure in Figure 1. In early stage experiements, longer sequences gave better validation auc score, so I wanted to find a way to input longer sequences. Through some experiments, I only used the last input of sequence $I_L$ as query vector in the transformer encoder instead of whole sequence from $I_1$ to $I_L$. The reason is that because I only predict the last question's correctness per input sequence, learning relationship between last question(query) and other questions(key) could be enough for model to work. As a result, it makes inner product QK in transformer encoder to have

O(L) time complexity, instead of O($L^2$). I found that performance is almost same with this speed advantage, at least when number of layers of transformer encoder is 1. I didn't use number of layers larger than 1 because using longer sequences gave better valid AUC correlation than using more layers. By this trick, I was able to input longer sequences, 5 models with L=1728 in final submission. Specific parameters for transformer encoder is d=128, number of layers=1, number of heads=2,4,8,16,32 for 5 different models. I used padding mask but no look ahead mask. One LSTM layer with d=128 is used over it.

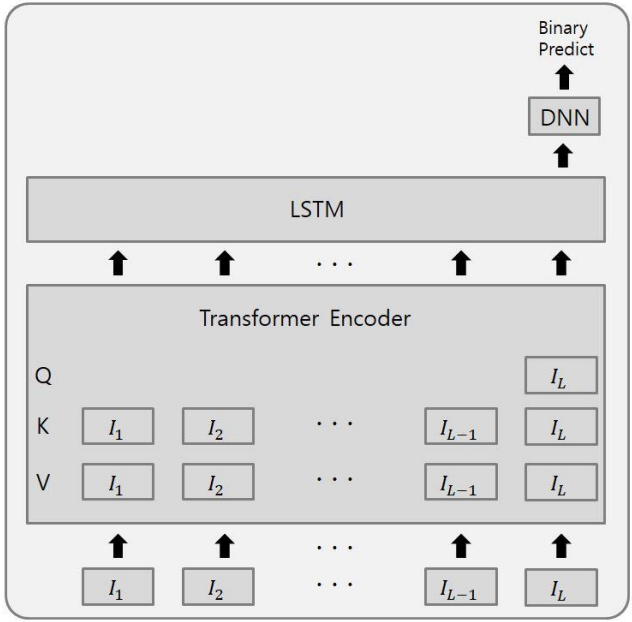

Figure 1. Model Structure

**Ensembling**

5 models with different transformer encoder heads 2,4,8,16,32 had similar validation AUC around 0.8165. Single model scored 0.816 AUC in public leaderboard, and 0.818 AUC in private. 5 models Ensemble scored 0.818 AUC in public leaderboard, and 0.820 AUC in private.

**Conclusion**

In this paper, I presented my winning model to the 'Riiid! Answer Correctness Prediction' competition. By combining transformer and RNN, I took the advantage of both model. By observing that using longer sequences give better results, I reduced the size of query vector from L to 1 in transformer encoder to change QK inner product to have O(L) time complexity instead of O($L^2$). This enabled me to win the competition by ensembling 5 models with input sequence length 1728 in final submission, which scored 0.820 AUC in private leaderboard.